\documentclass[12pt]{article}

\usepackage{graphicx,url}

\begin{document}

\title{Demographic differences in search engine use with implications for cohort selection}
\date{}

\maketitle

\author{
\centering
Elad Yom-Tov \\
Microsoft Research Israel \\
Herzeliya, Israel \\
eladyt@microsoft.com \\
}

\begin{abstract}

The correlation between the demographics of users and the text they write has been investigated through literary texts and, more recently, social media. However, differences pertaining to language use in search engines has not been thoroughly analyzed, especially for age and gender differences. Such differences are important especially due to the growing use of search engine data in the study of human health, where queries are used to identify patient populations.

Using data from multiple general-purpose Internet search engines gathered over a period of one month we investigate the correlation between demography (age, gender, and income) and the text of queries submitted to search engines. 

Our results show that females and younger people use longer queries. This difference is such that females make approximately 25\% more queries with 10 or more words. In the case of queries which identify users as having specific medical conditions we find that females make 50\% more queries than expected, and that this results in patient cohorts which are highly skewed in gender and age, compared to known gender balance.

Our results indicate that studies where demographic representation is important, such as in the study of health aspect of users or when search engines are evaluated for fairness, care should be taken in the selection of search engine data so as to create a representative dataset.

\end{abstract}

\section{Introduction}

Search engines are the single most widely used Internet service \cite{purcell2012}. For this reason, significant efforts have been dedicated to improve the interaction between search engines and their users, as evident from thousands of studies analyzing these interactions. Moreover, because of the popularity of search engines, their data has been used to study human behavior in areas ranging from politics \cite{diaz2016} to health \cite{yomtov2016}. 

One aspect which has received relatively little attention is the influence of user demographics, especially age and gender, on search engine use. This is surprising, since gender \cite{newman2008} and age \cite{pennebaker2003} are known to influence the use of language. Because of this, it might be assumed that language variation could be a useful for providing the best information to users, and could cause bias in the way results are provided to users.

Nevertheless, two seminal studies examined the correlation between demographics of search engine users and their topics of interest \cite{weber2010,weber2011}, finding that the topics people queried about varied by age and gender. Later, Bi \cite{bi2013} showed that search engine queries could be used to predict the age and gender of users. In a laboratory-based eye tracking study, Lorigo et al. \cite{lorigo2006} found gender differences in the way search engine results pages are analyzed. More recently Mehrotra et al. \cite{mehrotra2017} examined age and gender differences in the perception of search engine results, finding only minor differences therein.   

We note that in contrast to the dearth of literature on the effects of age and gender on search engine queries, a large body of work exists on detecting demographics of writers from their writings, e.g., for use in forensic analysis. A review of such methods is provided in Koppel et al. \cite{koppel2009computational}. Other work has focused on age and gender identification from social media \cite{schwartz2013,rangel2013}, as well as prediction of social class \cite{reoctiuc2015}. However, in contrast with both these lines of work, which rely on long text with complete sentences, search engine queries are (as shown below) short and often incomplete. 

The richness of search engine queries, reflecting a broad range of human behaviors, has led researchers analyze these data to learn about aspects of health which are difficult to study in other ways \cite{yomtov2016}. These studies begun from an examination of broad aspects of public health, e.g., the prevalence of influenza in a population \cite{polgreen2008} and the effect of dietary deficiencies on certain chronic pains \cite{giat2018}. More intriguingly, recent work has demonstrated that insights pertaining to individuals can be found their queries. These include, for example, the ability to identify precursors (including risk factors) for disease \cite{yomtov2015automatic}, and to screen for several forms of cancer, including pancreatic \cite{paparrizos2016}, ovarian and cervical \cite{soldaini2017}.         

In contrast to population-level analysis, where models are trained to predict area-level measures of health, studies of individual health require the identification of a group of people who share the medical condition under study. Identifying this group, also known as a cohort, is a challenging problem considering that search engine data is usually anonymous and rarely linked to medical information such as medical records. Thus, researchers have identified a group of users, called Self-Identified Users (SIUs) \cite{yomtov2015automatic}, who issue experiential queries \cite{paparrizos2016} such as ``I have ovarian cancer". SIUs were used either to identify the cohort \cite{paparrizos2016} or as a seed-set for algorithms which use these data in conjunction with other information to form the cohort. However, Soldaini and Yom-Tov \cite{soldaini2017} found that queries by SIUs differ substantially from those of other people they identified as sharing the same medical condition. This observation, if true, should have a dramatic influence on the representativeness of the cohort, especially if it is selected to include only SIUs.   

Thus, in this work we seek to examine how user's demographics influence their choice of queries. The differences we find could affect the quality of results that search engines return, thus having an important effect on the fairness of the results served by search engines, and are consequential to studies of cohorts based on SIUs. 

\section{Methods}

Three main datasets from two sources were used in this study. First, a sample of approximately 5.5 million queries submitted to the Bing search engine from users in the USA during one day, 21st March 2018. For each query we obtained the text of the query, the age group and gender (male or female) of the user. The latter two were as reported by users during registration to Bing. We refer to these as dataset 1. 

To validate our findings from the analysis of the first dataset, we performed identical analyses on 5 weeks of query data collected from an opt-in consumer panel recruited by Internet analytics company comScore. Millions of panelists provide comScore with explicit permission to passively measure all of their online activities using monitoring software installed on their computers. In addition to logged search behavior, the comScore data also collects panelists' gender and age group. This dataset contains queries made not only to Bing. We refer to these as dataset 2. 

Finally, we collected all experiential queries submitted to Bing during March 2018. These queries consisted of all queries containing the phrase ``been diagnosed with $<$condition$>$" or ``I have $<$condition$>$", without queries that only indicate possibilities (``Do I have $<$condition$>$") or negation (``I have not been diagnosed with $<$condition$>$"). Conditions were one of 5521 conditions and their 25,584 synonyms, as used in \cite{yomtov2015automatic}. Here again, for each query we obtained the text of the query, the age group and gender (male or female) of the user. 

To evaluate spelling mistakes in the text of queries we used Python's Language Check package. In order to evaluate queries for whether they formed complete sentence(s) we randomly sampled 50 queries from each age group and gender in dataset 1. These queries were labeled by 5 crowdsourced workers on the Crowdflower platform as to whether they were complete sentences. We analyzed all those queries which had an agreement of 4 or more workers.  

\section{Results}

\subsection{General queries}

\begin{figure}[t]
  \centering 
    \includegraphics[width=0.75\textwidth]{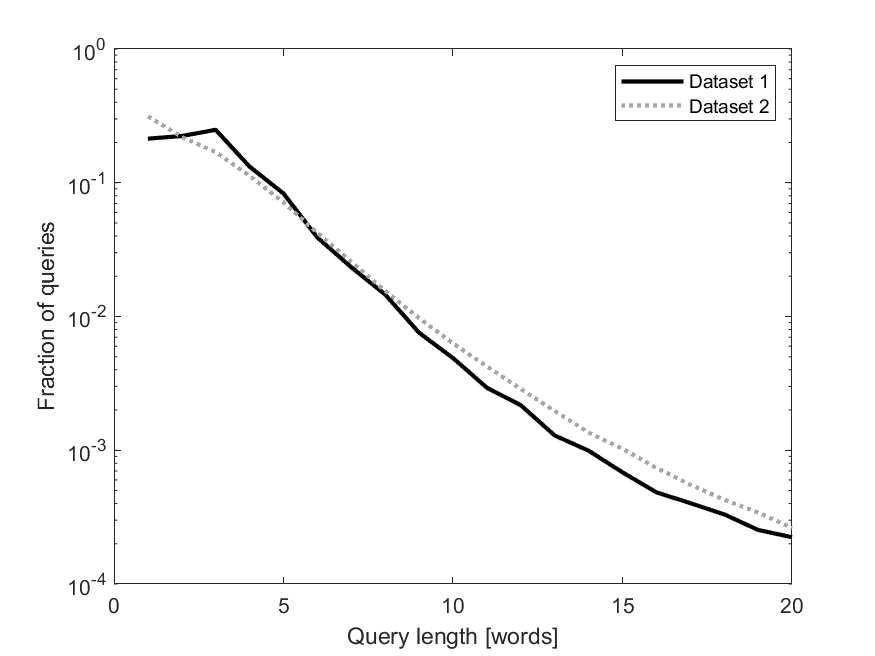}
  \caption{Fraction of queries of each length. The vertical axis is log-scaled.}
    \label{fig:length}  
\end{figure}

The average number of words per query was 3.2 in dataset 1 and 3.0 in dataset 2. This is in agreement with previously reported average lengths, e.g., \cite{song2013}. The distribution of queries as a function of query length is shown in Figure \ref{fig:length}. As the figure shows, the distribution is highly skewed to shorter queries, and is similar in both datasets. 

\begin{figure}[t]
  \centering 
    \includegraphics[width=0.75\textwidth]{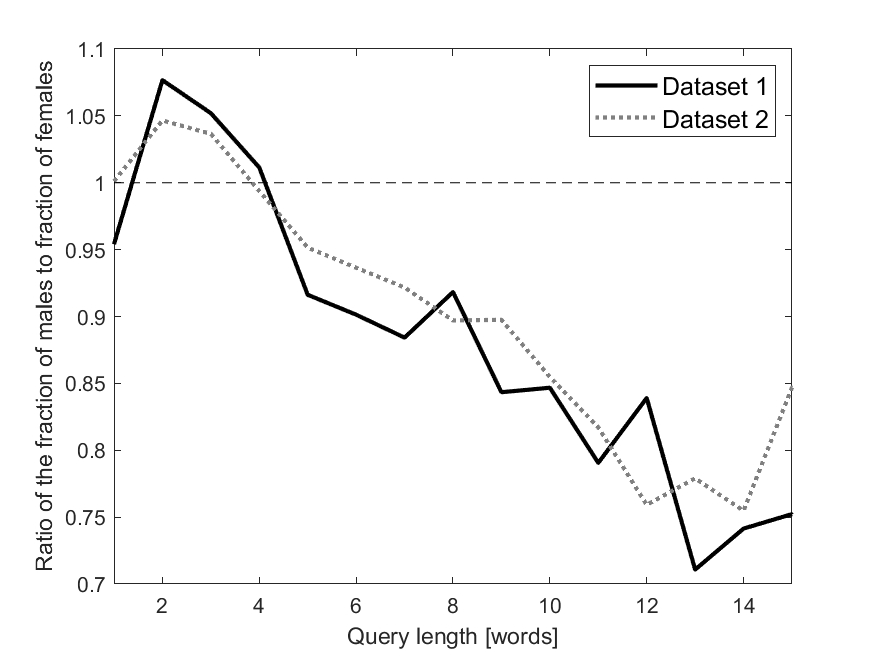}
  \caption{Ratio between the fraction of queries at each length made by men to the fraction of queries at this length made by women}
    \label{fig:gender}  
\end{figure}

We computed the fraction of queries at each length, stratified by age group and gender. Figure \ref{fig:gender} shows the ratio between the fraction of queries of length $N$ made by males, compared to the fraction as made by females, for both datasets. As the figure shows, the datasets are extremely similar to each other. Interestingly, queries with 2-4 words are approximately 5\% more common among males, but longer queries are much more common among females, with a clear correlation between query length and preference by females. 

\begin{figure}[t]
  \centering 
    \includegraphics[width=0.75\textwidth]{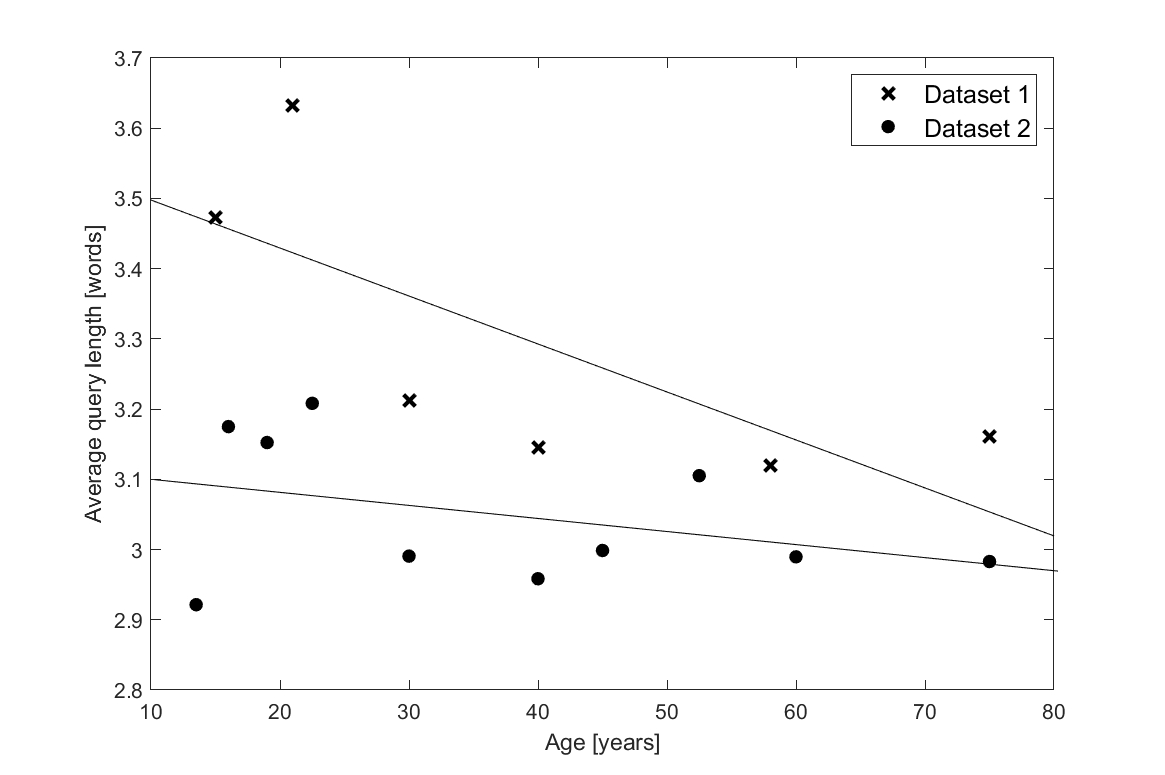}
  \caption{Average number of words by age. Lines are linear regression lines, the top for dataset 2 and the bottom for dataset 1. }
    \label{fig:age}  
\end{figure}

Figure \ref{fig:age} shows the average query length for each age group. Each age group is represented by the middle of the age group (i.e., the age group of 18-20 year olds is represented by a point at 19 years). As the figure shows, younger people use longer queries, though the effect is small (between 3.6 words to 2.9 words). 

Income level (provided in 7 groups for dataset 2) is uncorrelated with query length.

The average number of spelling mistakes per word per age group is correlated with age (adjusted $R^2$ of $0.88$, $p=0.005$), with higher aged users making fewer spelling mistakes (0.296 per word for the youngest age group and 0.230 for the oldest ones). The difference between males (0.266) and females (0.260) is negligent.

Finally, we evaluated whether the longer queries (by females and younger users) were due to these users making queries that were complete sentences. As noted above, crowdsourced workers labeled a sample of the queries for whether they formed complete sentences (e.g., ``Do I have the flu?", vs. ``flu"). A multi-way ANOVA with interactions found that gender was not statistically significantly associated with the use of complete sentences, while age ($P=0.007$) and the interaction of age and gender ($P=0.010$) were. Younger users were more likely to query using complete sentences.

\subsection{Experiential queries}

In the previous section it was shown that females use longer queries than males. Experiential queries of the type  previously used to identify cohorts for study, or as a seed-set to find cohorts using auxiliary information, are naturally at least 3 words in length (e.g., ``I have cancer") and usually far longer (e.g., ``I was diagnosed with stage 2 lung cancer"). Therefore, here we examine how the demographic bias in query length is reflected in experiential queries.

As described in the Methods, we extracted experiential queries from one month of Bing data.

The 10 most common conditions mentioned in these queries, in descending order of popularity, were: cancer, allergy, hernia, yeast infection, cyst, pimple, bleeding, pregnancy, black eye, blister. 

The fraction of experiential queries made by males was found to be 36.2\% lower than expected, according to their fraction in the population. Conversely, the fraction of experiential queries made by females was 50.0\% greater than expected. This is to be expected, since the average length of experiential queries was 13.2 words, and (as shown above) females make longer queries than males. 

\begin{figure}[t]
  \centering 
    \includegraphics[width=0.75\textwidth]{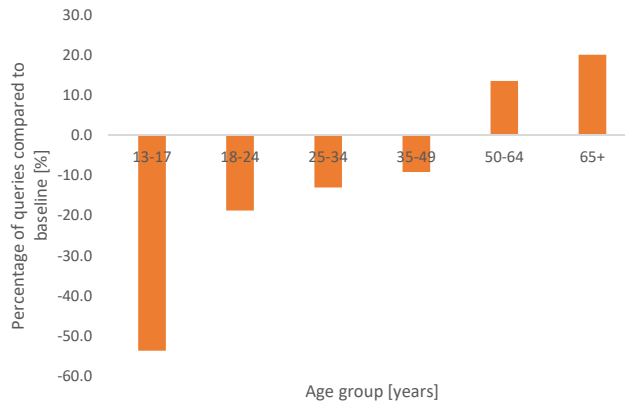}
  \caption{Relative propensity of experiential queries, compared to the baseline (all queries), by age group.}
    \label{fig:sius}  
\end{figure}

Figure \ref{fig:sius} shows the fraction of experiential queries by age group, compared to the baseline of all queries submitted to Bing. The figure shows a correlation between age and the excess in experiential queries, perhaps partly because the incidence of diseases is higher among older people. 

Finally, in the 18 cancers for which at least 20 people made  experiential queries, we compared the ratio of males to females who made these queries, compared to known gender ratio\footnote{As provided by either Cancer Research UK (\url{www.cancerresearchuk.org}), American Cancer Society (\url{cancer.org}) or published in the scientific literature.}. To model the relationship between the known gender ratio and the observed gender ratio (on Bing), we used a linear model where the independent variable was the known gender ratio and the independent variable the observed ratio. The model reached an $R^2$ of 0.35, with a slope of 0.5. This means that the gender ratio in experiential queries is strongly biased towards females, as expected by the above results. 

\section{Discussion}

Our analysis of search engine logs from two sources reveal that search queries differ by the demographics of those querying. Specifically, age and gender (but not income) have a statistically significant effect on query length, where females and younger people appear to make more longer queries. The findings are in agreement with studies of longer texts \cite{koppel2009computational}, which found similar differences among ages and genders.

The similarity among datasets shows that the difference is not because of the demographics of a specific search engine (e.g., Bing). This is in agreement with past studies which did not find a difference in the demographics of search engine users \cite{yomtov2018}.

Our findings have two areas of impact. First, they contribute to the knowledge on the differences in search engine use among people, and the need to take these into account when measuring issues such as fairness and bias in search engine results.

Second, our results have a clear implication for studies which use anonymous query data to study aspects of human health. In such studies a cohort needs first to be identified. Several methods for this have been suggested, but all rely on experiential queries, either as the sole source of data or as a supporting source. Since our work has shown that experiential queries are highly skewed by gender and age, by implication, so is the cohort, especially when based only on experiential queries. Future work is required to assess if any of the previously suggested methods for cohort analysis \cite{yomtov2015automatic,ofran2012,soldaini2017} also suffer from this imbalance, or if their use of additional data mitigates this issue.

\bibliographystyle{plain}
%\bibliography{bibliography} 

\end{document}